# Wikipedia Cultural Diversity Dataset:

# A Complete Cartography for 300 Language Editions

**Marc Miquel-Ribé**

Universitat Pompeu Fabra, Catalunya
marcmiquel@gmail.com

**David Laniado**

Eurecat, Centre Tecnològic de Catalunya
david.laniado@eurecat.org

**Abstract**

In this paper we present the Wikipedia Cultural Diversity dataset. For each existing Wikipedia language edition, the dataset contains a classification of the articles that represent its associated cultural context, i.e. all concepts and entities related to the language and to the territories where it is spoken. We describe the methodology we employed to classify articles, and the rich set of features that we defined to feed the classifier, and that are released as part of the dataset. We present several purposes for which we envision the use of this dataset, including detecting, measuring and countering content gaps in the Wikipedia project, and encouraging cross-cultural research in the field of digital humanities.

## Introduction

By making all its content and interactions available, the online encyclopaedia Wikipedia has become a "living laboratory", ideal for empirical studies (Schroeder & Taylor, 2015). There is abundant scholarly research on how editors collaboratively create the articles, content quality, and the final consumption by its readers (Lemmerich, Sáez-Trumper, West & Zia, 2018; Mesgari, Okoli, Mehdi, Nielsen, & Lanamäki, 2015; Okoli, 2014; Okoli, Mehdi, Mesgari, Nielsen, & Lanamäki, 2012). Nonetheless, most of the studies are based on the English language edition, neglecting the fact that the project exists in 301 language editions, and the subsequent diversity both in terms of editors' organization and content topics.

Wikipedia language editions present different arrays of topics from each other to the point that there is only a partial overlap between bigger language editions (e.g. English and German) and among those geographically close to each other (Hecht & Gergle, 2010b; Warncke-Wang, Uduwage, Dong, & Riedl, 2012). This lack of correspondence of content between language editions has been named 'language gap' and implies that the creation of content by each community obeys to different dynamics influenced by cultural and contextual factors (Hecht & Gergle, 2010a; Samoilenko, Karimi, Edler, Kunegis, & Strohmaier, 2016).

In order to understand better the composition of each Wikipedia language edition, in our previous work (Miquel-Ribé, 2016; Miquel-Ribé & Laniado, 2016) we developed a methodology to identify the articles related to the editors' geographical and cultural context (i.e. their places, traditions, language, agriculture, biographies, etc.). We named such articles **Cultural Context Content (CCC)**. Results showed that among the largest 40 Wikipedia language editions, these articles represent on average the 25% of all the content and are mostly exclusive, i.e. they have no equivalence across language editions. This confirms that a large part of the language gap is due to cultural and geographical specificities, i.e. there exists a culture gap which limits cultural diversity within each language edition's content.

It is easy to argue that, for each language edition to achieve Wikipedia's goal of "gathering the sum of human knowledge", editors should necessarily consider all the different points of views implicit in the content of the rest of language editions. Yet, studies on multilingualism activities in Wikipedia show the difficulties of content exchanges across languages, which mostly happen with incursions to the English language edition made by a minority of very participative editors (Hale, 2014; Kim et al., 2016).

Bearing this in mind, we created the project ***Wikipedia Cultural Diversity Observatory***[1] as a space for both scholars and editors to study Wikipedia intercultural coverage and counter content gaps. Thus, the project aims to raise awareness on Wikipedia's current state of cultural diversity by providing datasets, visualizations and statistics, as well as pointing out solutions and tools.

---

[1] http://wcdo.wmflabs.org

In this paper we present a complete dataset created through an improved version of the methodology to obtain CCC articles, applied to all the existing Wikipedia language editions. The method was improved by introducing external data from Wikidata, and using a machine learning classifying technique. Additionally, in order to verify the quality of the final classification, we repeated the manual assessment which showed an improvement in relation to previous results.

The dataset has a record for each article from each language edition, containing general features of the article and its history, such as its number of incoming links and number of edits, and a rich set of features describing its relation to the cultural context, such as geo-coordinates assigned to it, territory names or demonyms contained in its title, or in categories to which it belongs, semantic properties such as *place of birth* in case of a biography, or *original language* in case of literary work.

The Wikipedia Cultural Diversity dataset has several applications which can be divided into: a) further analysis of the culture gap and article suggestions in order to bridge it, b) research in the field of digital humanities, and c) the use of contextual data to feed automatic applications.

## Dataset Creation

### Language-Territories Mapping

Obtaining a collection of Cultural Context Content (CCC) for a language requires associating each language to a list of territories, in order to collect everything related to them as a context. We chose to consider as territories associated to a language the ones where that language is spoken as native indigenous or where it has reached the status of official. We selected the political divisions of first and second level (this is countries and recognized regions). Many languages could be associated to countries, i.e. first level divisions, and second level divisions were used only when a language is spoken in specific regions of a country.

In order to identify such territories, we used ISO codes. First and second level divisions correspond to the ISO 3166 and ISO 3166-2 codes. These codes are widely used on the Internet as an established standard for geolocation purposes. For instance, Catalan is spoken as an official language in Andorra (AD), and in Spain regions of Catalonia (ES-CA), Valencia (ES-PV) and Balearic Islands (ES-IB). For the Italian Wikipedia, the CCC comprises all the topics related to the territories (see dark blue in the map): Italy (IT), Vaticano (VA), San Marino (SM), Canton Ticino (CH-TI), Istria county (HR-18), Pirano (SI-090) and Isola (SI-040), whereas, for the Czech language, it only contains Czech Republic (CZ). A widespread language like it is English comprises 90 territories, considering all the countries where it is native and the ex-colonies where it remains as an official language, which implies that the CCC is composed by several different contexts.

The language territories mappings compound a database[2] with the ISO code, the Wikidata qitem and some related words for each territory. In particular, we include the native words to denominate each territory, their inhabitants' demonyms and the language names (e.g., eswiki españa mexico … español castellano). This word list has been initially generated by automatically crossing language ISO codes, Wikidata, Unicode and the Ethnologue databases, which contain the territories where a language is spoken and their names in the corresponding language. The generated list for each territory has been subsequently manually revised and extended (using information from the specific articles in the correspondent Wikipedia language edition). Wikipedians were invited to suggest changes and corrected a few lines of the database (e.g. regions where Ukranian is spoken in countries surrounding Ukrania).

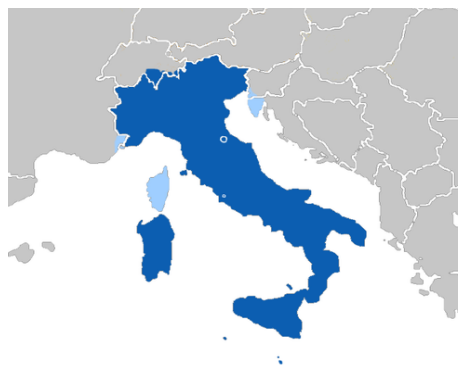

*Figure 1. In dark blue the territories where Italian is spoken natively. In light blue where it is used as a secondary language.*

### Feature Description

Once we obtained the language-territory mapping database, we defined a set of features representing the association between an article and a language.
In order to access the Wikipedia data, we used the MySQL replicas (real-time exact copies of the databases) of each language edition provided by the Wikimedia Foundation. We complemented it with data extracted from the Wikidata XML dumps generated on a monthly basis. Wikidata is a secondary database used by Wikipedia, created in the same collaborative fashion, where every article is linked to a qitem, an entity which includes several properties that describe it[3].

---

[2] The mapping tables can be seen on GitHub: https://github.com/marcmiquel/WCDO/tree/master/language_territories_mapping

The method to obtain CCC integrates several features described below in order to qualify all the Wikipedia language edition articles as (1) *reliably CCC*, (2) *potentially CCC*, (3) *reliably non CCC* and (4) *potentially non CCC*. These labels express the degree of certainty with which we consider an article should belong to CCC, according to some basic features. For example, we established that articles associated to a territory through geocoordinates or through specific keywords in their titles such as the territory name or the demonym should reliably be included into CCC collection.

These features were fed into a machine learning classifier and included in the final dataset. Features 1, 2, 3 and 12 were already used in (Miquel-Ribé & Laniado 2018), where they are described in further detail.

### Feature 1: Geolocated articles (*reliably CCC*)
This first feature is derived from the geocoordinates and the ISO code found in Wikidata and the mysql geotags table. As the usage of geocoordinates and ISO codes is not uniform across language editions and may contain errors, a reverse geocoder tool was used to check the ISO 3166-2 code of the territory each geolocated article.

### Feature 2: Keywords on title (*reliably CCC*)
The second Feature was obtained looking at article titles and checking whether they contain keywords related to a language or to the corresponding territories (e.g., "Netherlands National football team", "List of Dutch writers", etc.).

### Feature 3: Category crawling (*potentially CCC*)
The third feature was derived from the category graph. Each article in Wikipedia is assigned directly to some categories, and categories can in turn be assigned to higher level categories. We then started from the same list of keywords used for feature 2, and identified all the categories including such keywords. For example, "Italian cheeses" or "Italian cuisine". We then took all articles contained in these categories, and iteratively went down the tree retrieving all their subcategories and the articles assigned to them. In this way we did not only get a binary value, but also discrete indicators for an article: the shortest distance in the tree from a category containing a relevant keyword, and the number of paths connecting the article to one of such categories. As the category trees may be noisy, we did not consider this feature reliable, and we assigned the articles retrieved in this way to the group of *potentially CCC* articles.

### Wikidata properties
Wikidata properties were used as additional features to qualify articles. Every article corresponds to one entity in Wikidata identified by a qitem, and has properties whose values correspond to the qitems of other entities. Such entities might in turn correspond to the language or to the territories associated to it, bringing valuable information for our aim. Hence, we created several groups of properties and qualified each article in order to ascertain whether it is reliably or potentially part of CCC.

### Feature 4: Country properties (*reliably CCC*)
P17 (country), P27 (country of citizenship), P495 (country of origin) and P1532 (country for sport).

Entities for which some of these properties refer to countries mapped to the language, as established in the language-territories mapping, are directly qualified as reliably CCC. These entities are often places or people.

### Feature 5: Location properties (*reliably CCC*)
P276 (location), P131 (located in the administrative territorial entity), P1376 (capital of), P669 (located on street), P2825 (via), P609 (terminus location), P1001 (applies to jurisdiction), P3842 (located in present-day administrative territorial entity), P3018 (located in protected area), P115 (home venue), P485 (archives at), P291 (place of publication), P840 (narrative location), P1444 (destination point), P1071 (location of final assembly), P740 (location of formation), P159 (headquarters location) and P2541 (operating area).

Entities for which some of these properties have as value a territory mapped to the language are directly qualified as reliably part of CCC. Most usually, these properties have as values cities or other more specific places. Hence, the method employed uses in first place the territories from the Languages Territories Mapping in order to obtain a first group of items, and next it iterates several times to crawl down to more specific geographic entities (regions, subregions, cities, towns, etc.). Therefore, all articles were finally qualified as located in a territory or in any of its contained places. It is good to remark that not all of the location properties imply the same relationship strength.

### Feature 6: Strong language properties (*reliably CCC*)
P37 (official language), P364 (original language of work) and P103 (native language).

Entities associated through some of these properties with the qitem of the language (or of one of its dialects) were directly qualified as reliably part of CCC. This property was used both for characterizing works (from theatre plays to monuments) and people.

---

[3] https://www.wikidata.org/wiki/Wikidata:Introduction

**Feature 7: Created_by properties (*reliably CCC*)**

P19 (place of birth), P112 (founded by), P170 (creator), P84 (architect), P50 (author), P178 (developer), P943 (programmer), P676 (lyrics by) and P86 (composer).

Entities associated through some of these properties with one of the entities already qualified as reliably CCC are also directly qualified as reliably part of CCC. Although some of these relationships can be fortuitous, we considered them as important enough in order to qualify one article as CCC, assuming a broader interpretation of which entities are involved in a cultural context. This property is usually used for characterizing people and works.

**Feature 8: Part_of properties (*reliably CCC*)**

P361 (part of).

Entities associated through this property with one of the entities already qualified as reliable CCC were also directly qualified as part of CCC. This property is used mainly for characterizing groups, places and work collections.

**Feature 9: Weak language properties (*potentially CCC*)**

P407 (language of work or name), P1412 (language spoken) and P2936 (language used).

These properties are related to a language but present a weaker relationship with it. Therefore, entities associated through some of these properties with the language (or one if its dialects) may be related to it in a tangential. Hence, they were qualified as potentially CCC.

**Feature 10: Affiliation properties (*potentially CCC*)**

P463 (member of), P102 (member of political party), P54 (member of sports team), P69 (educated at), P108 (employer), P39 (position held), P937 (work location), P1027 (conferred by), P166 (award received), P118 (league), P611 (religious order), P1416 (affiliation) and P551 (residence).

Entities associated through some of these properties with one of the entities already qualified as reliably CCC are potentially part of CCC. Affiliation properties represent a weaker relationship than created_by. It is not possible to assess how central this property is in the entities exhibiting it, hence these were qualified as potentially CCC.

**Feature 11: Has_part properties (*potentially CCC*)**

P527 (has part) and P150 (contains administrative territorial entity).

Entities associated through some of these properties with one of the entities already qualified as reliably CCC are potentially part of CCC, as they could be bigger instances of the territory that might include other territories outside the language context.

**Feature 12: Inlinks from / Outlinks to CCC groundtruth (*potentially CCC*)**

This feature aims at qualifying articles according to their incoming and outgoing links, starting from the assumption that concepts related to the same cultural context are more likely to be linked to one another. Hence, for each article we counted the number of links coming from other articles already qualified as reliably CCC (*inlinks from CCC*), and computed the percentage in relation to all the incoming links (*percent of inlinks from CCC*) as a proxy for relatedness to CCC.

Likewise, for each article we counted the number of links pointing to other articles already qualified as reliably CCC (*outlinks to CCC*) and the corresponding percentage with respect to their total number of outlinks (*percent of outlinks to CCC*). We expect a high percentage of outlinks to CCC to imply that an article is very likely to be part of CCC, as its content refers to that cultural context.

**Other Languages CCC Features**

**Feature 13: Geolocated articles not in CCC (*reliably non CCC*)**

Articles that are geolocated in territories associated to other languages are directly excluded from being part of a language's CCC. Even though there might be some exceptions, articles geolocated out of the territories specified in the language-territory mapping for a language are reliably part of some other language CCC.

**Features 14 and 15: Location not in CCC property (*reliably non CCC*) and Country not in CCC property (*reliably non CCC*)**

For the Wikidata properties country_wd and location_wd presented above, we checked whether they referred to territories not associated to the language. Hence, similarly to the previous feature, they are reliably related to some other language CCC.

**Features 16: Inlinks / Outlinks to geolocated articles not in CCC (*potentially non CCC*)**

The last feature aims at qualifying articles according to how many of their links relate to territories which are not mapped to the language. Similarly to Feature 12, the number of inlinks and outlinks to geolocated articles not mapped to the language were counted along with their percentual equivalent (i.e. *inlinks from geolocated not in*

*CCC, percent inlinks from geolocated not in CCC, outlinks to geolocated not in CCC, percent outlinks to geolocated not in CCC*). Articles qualified by these features are potentially part of other languages CCC.

## Machine Learning

The above described features were used to qualify all the articles from each Wikipedia language edition and to feed a classifier in order to expand the *reliably CCC* set collected up to this point. The scikit implementation[4] of the machine learning classifier Random Forest was used, with 100 estimators.

Before training the classifier, we assigned class 1 to the articles whose features were qualified as *reliably CCC*, and class 0 to the articles whose features were reliably non CCC. A few articles had both kinds of features coexisting, but these were a tiny minority and, in this way, we could ensure that there would be no undesired articles in class 1 - the final selection.

To train the classifier, we introduced the positive group (class 1). Since the negative group (class 0) was composed mainly by articles with geolocation and territory-based properties, we considered that they were not representative enough of the entire group. Hence, we decided not to use them as the negative group for training the classifier (class 0). Instead, we employed a negative sampling process (Dyer, 2014), in which all the articles not in class 1 were retrieved and introduced 5 times as class 0, even though they included unqualified articles, articles qualified as potentially CCC, articles qualified as potentially non CCC and articles qualified as a *reliably non CCC*. In other words, the classifier was trained to distinguish positive articles from random articles.

Finally, the classifier was fed with the fitting data which needed to be categorized as class 1 or class 0. The data introduced were all the *potentially CCC* articles. We used a machine learning classifier based on a multiple path algorithm in order to calculate the weight of each feature to determine whether an article belongs to class 1 or 0.
The accuracy provided estimated by the classifier is in the order of 0.999, and some features like the percentage of outlinks to CCC, percentage of outlinks to other CCC and category crawling level emerged as particularly relevant.

## Manual Assessment

Manual assessment was performed to test the accuracy of the classifier. The same process was followed as in previous work (Miquel-Ribé and Laniado, 2018), which was used as a baseline.

The Japanese and German Wikipedia editions were used to compare the results obtained by the algorithm with the ones manually assigned by three human raters and to calculate the Cohen's Kappa coefficient (Cohen 2016). The expert raters accessed the content of each article and classified it as belonging to CCC or not. Results, reported in Table 1, show that the degree of agreement with expert judgement is generally higher than for the baseline, getting to be in some cases comparable to the agreement between human raters.

*Table 1. Cohen's Kappa coefficients for the Japanese and German Wikipedia editions. Agreement between the results obtained by the algorithm and by human raters in the Wikipedia Cultural Diversity dataset. Coincidence (coinc.) is the proportion of agreement, and K is the Cohen's Kappa coefficient. Results from (Miquel-Ribé and Laniado, 2018) are reported as a baseline. Inter-rater agreement between the three human raters is also reported.*

|  | Japanese | | German | |
|---|---|---|---|---|
| **Results** | coinc. | K | coinc. | K |
| algorithm-rater1 | 0.95 | 0.90 | 0.96 | 0.91 |
| algorithm-rater2 | 0.93 | 0.86 | 0.96 | 0.92 |
| algorithm-rater3 | 0.93 | 0.87 | 0.87 | 0.87 |
| **Baseline** | coinc. | K | coinc. | K |
| algorithm-rater1 | 0.86 | 0.71 | 0.90 | 0.80 |
| algorithm-rater2 | 0.89 | 0.77 | 0.91 | 0.82 |
| algorithm-rater3 | 0.86 | 0.72 | 0.89 | 0.77 |
| **Inter-rater** | coinc. | K | coinc. | K |
| rater1-rater2 | 0.97 | 0.94 | 0.96 | 0.93 |
| rater1-rater3 | 0.97 | 0.93 | 0.95 | 0.90 |
| rater2-rater3 | 0.96 | 0.91 | 0.98 | 0.95 |

We then repeated the manual assessment procedure with one human rater for a larger sample of language editions. A sample of 10 language editions was created, picking languages in such way to maximize diversity in terms of size of the encyclopedia, geographical spread and location. For each of these 10 languages we randomly picked for manual evaluation 100 articles classified by the algorithm as positive (belonging to CCC) and 100 articles classified as negative (not belonging to CCC). The results are presented in Table 2, which details the percentage of false positives (FP) and false negatives (FN) with the resulting F1 score for each language edition. False positives are on average the 2.7%, false negatives the 3.3%. The average value of F1 is 0.97. These results indicate a clear tendency to improvement with respect to the ones from (Miquel and Laniado 2017), where false positives were reported to be on average the 8.1% and false negatives the 5.9%, with an average F1-score of 0.92.

---
[4] https://scikit-learn.org/stable/modules/generated/sklearn.ensemble.RandomForestClassifier.html

*Table 2. Results of the manual assessment. For each language the total number of articles (Articles) and the proportion of articles classified as CCC (CCC%) are reported, together with the percentages of False Positives (FP %) and False Negatives (FN %), and the resulting F1-score (F1) .*

| ISO Code | Articles | CCC % | FP % | FN % | F1 |
|---|---|---|---|---|---|
| ca | 584,760 | 17.1% | 2 | 4 | 0.98 |
| de | 2,195,308 | 33.7% | 1 | 2 | 0.99 |
| en | 5,676,573 | 44.2% | 5 | 5 | 0.95 |
| fa | 629,125 | 21.9% | 6 | 1 | 0.94 |
| gn | 715 | 19.9% | 3 | 6 | 0.97 |
| ja | 1,110,617 | 51.0% | 1 | 4 | 0.99 |
| ms | 306,055 | 22.1% | 1 | 0 | 0.99 |
| ru | 1,481,560 | 32.2% | 0 | 3 | 1.00 |
| sw | 42,422 | 19.0% | 7 | 5 | 0.93 |
| zu | 1,111 | 14.22% | 1 | 3 | 0.99 |

### *Main_territory* Attribution

The proportion of articles included in CCC over the 300 language editions ranges from 74.8% (Muscogee) to 0.04% (Waray), with an average of 15.58% and a median of 11.91%. Considering the 25 biggest language editions, the average is 20.73% and the median 21.77%.

A set of CCC articles is defined for each language edition. However, as previously explained, it contains articles which can relate to the different territories mapped to a language. This aspect has special importance when a language is spread across different countries and even continents.

In order to quantify how many articles can be attributed to each specific territory, we created a simple heuristic to estimate a value for the column named 'main territory' for each article:

a) Geolocated articles can be easily attributed to a main territory (first level or second level) using ISO-3166 and ISO-3166-2 correspondingly.
b) Articles which contain a keyword in their title such as the territory name or demonym (and it is not the language name) can be attributed to that territory.

For the rest of CCC articles, we considered a rule of majority in order to associate a language to a territory. We considered the columns category_crawling_territories, country_wd, location_wd and checked the qitems they contained and counted their number of occurrences. An article is associated to the territory whose number of occurrences is higher. In case of tie values, the article characteristic main territory remains Unassigned.

In case an article contains any qitem in these columns (part_of_wd, has_part_wd, created_by_wd and affiliation_wd), this is not a territory qitem but another CCC article's qitem. In these cases, only when these CCC articles were associated to a main territory it is possible to use their value in order to count the number of occurrences and assign the main territory.

Once CCC articles are assigned to a main territory we can see their distribution among the different territories mapped to a language. For instance, results for the German CCC present the following ranking: Germany (79.21%), Austria (10.94%), Switzerland (6.69%), Unassigned (2.15%), Luxembourg (0.37%), Silesian (0.17%), Ústí nad Labem (0.13%), Liechtenstein (0.11%), among others. Instead, results for the Italian CCC show a distribution with Italy (94.43%), Unassigned (3.72%), Ticino (0.51%), Izola (0.41%), San Marino (0.4%), Graubünden (0.28%), Vatican City (0.17%), Istria (0.07%) and Piran (0.01%).

The main_territory field provides higher resolution to the dataset and enhances its applications, as it allows to filter the content for each of the territories which compound the context for a given language context.

## Dataset Description

The dataset includes a file per language. The CSV format was chosen to facilitate further processing. Files are compressed using bzip2. The biggest file is the English Wikipedia (265MB) and the entire dataset is 1.67GB.

The dataset is available on Figshare[5] and at the Wikipedia Cultural Diversity Observatory dataset server[6]. It is also available in the form of a single SQLite 3 database[7] for all languages (named as ccc_old.db), which occupies 9.5GB.

**Dataset Structure**

There is a CSV file for each language edition including all the articles from the CCC collection. Each file contains one article per line with the following 52 columns:

- **general data columns:** qitem (from Wikidata), pageid (in the local Wikipedia), page title, date created (creation date timestamp), geocoordinates, ISO 3166 and ISO 3166-2.

- **ccc columns:** ccc binary (1 when the article belongs to CCC, 0 when it does not), main territory (qitem of the territory the article relates to) and number of retrieval strategies (number of different types of relationships to CCC, reliable or potential, whether they are geocoordinates, category crawling, etc.).

- **reliably ccc features:** ccc geolocated (1 when it is part of CCC, -1 when it belongs to another language's CCC), country wdproperties (property:qitem of the country it relates to), location wdproperties (property id and qitem), language strong wdproperties (property id and qitem), created by wdproperties (property id and qitem), part of wdproperties (property id and qitem) and keyword on title (qitem associated to the territory or language name).

- **potentially ccc features:** category crawling territories (territory qitem from which this article was found through category crawling), category crawling level (category graph level where the article has been found), language weak wdproperties (property id and qitem), affiliation wdproperties (property id and qitem), has part wdproperties (property id and qitem), number of inlinks from CCC (number of incoming links to the article from articles having some reliable CCC features, such as geolocated articles), number of outlinks to CCC (number of outgoing links from the article to articles having some reliable CCC features, such as geolocated articles), percent inlinks from CCC (number of such incoming links normalized with respect to all the incoming links), percent outlinks to CCC (number of such outgoing links normalized with respect to all the outgoing links).

- **reliably non CCC features:** other ccc country wdproperties (property id and qitem) and other ccc location wdproperties (property id and qitem).

- **potentially non CCC features:** other ccc language strong wdproperties (property id and qitem), other ccc created by wdproperties (property id and qitem), other ccc part of wdproperties (property id and qitem), other ccc language weak wdproperties (property id and qitem), other ccc affiliation wdproperties (property id and qitem), other ccc has part wdproperties (property id and qitem), number of inlinks from geolocated abroad (number of incoming links to the article from those articles that reliably associate to another CCC features such as geolocated articles in other CCC), number of outlinks to geolocated abroad (number of outgoing links from the article to those articles which reliably associate to another language CCC such as geolocated articles in other CCC), percent inlinks from geolocated abroad (number of such incoming links divided by all the incoming links) and percent outlinks to geolocated abroad (number of such outgoing links divided by all the outgoing links).

- **relevance features:** number of inlinks, number of outlinks, number of bytes, number of references, number of edits, number of editors, number of edits in discussions, number of pageviews (during the last month), number of wikidata properties and number of interwiki links, and featured_article (1 or 0 whether it is or not).

As an example, to illustrate the dataset, in Table 3 we present a record from the Italian Wikipedia for article "Parmigiano Reggiano" with its corresponding fields. The article represents a typical Italian cheese, and is labelled as part of CCC for the Italian Wikipeda (attribute "ccc_binary" is 1). We can see that it is indeed associated to the Italian cultural context through various features, such as properties "Country of origin" and "Location of final assembly", pointing to "Emilia Romagna", an Italian region. The article is strongly integrated in the Italian cultural context, with 86.5% of incoming links and 27.8% of outgoing links connecting it to CCC articles.

---

[5] https://doi.org/10.6084/m9.figshare.7039514.v3

[6] https://wcdo.wmflabs.org/datasets

[7] https://wcdo.wmflabs.org/databases

*Table 3. An example record of the dataset from the Italian Wikipedia (article "Parmigiano Reggiano"). The names in English of Wikidata properties and entities (qitems) are reported in parentheses.*

| Feature | value |
|---|---|
| qitem | Q155922 |
| page_title | Parmigiano_Reggiano |
| date_created | 20040913 |
| geocoordinates | |
| iso3166 | |
| iso31662 | |
| ccc_binary | 1 |
| main_territory | Q38 (Italy) |
| num_retrieval_strategies | 5 |
| country_wd | P495:Q38 (country of origin: Italy) |
| location_wd | P1071: Q1263: Q38; P1071: Q16228: Q38 (location of final assembly: Emilia-Romagna: Italy; location of final assembly: Province of Parma) |
| language_strong_wd | |
| created_by_wd | |
| part_of_wd | |
| keyword_title | |
| category_crawling_territories | Q38;Q652 (Italy;Italian) |
| category_crawling_level | 1 |
| language_weak_wd | |
| affiliation_wd | |
| has_part_wd | |
| num_inlinks_from_CCC | 122 |
| num_outlinks_to_CCC | 206 |
| percent_inlinks_from_CCC | 0.865 |
| percent_outlinks_to_CCC | 0.278 |
| other_ccc_country_wd | |
| other_ccc_location_wd | |
| num_inlinks_from_geolocated_abroad | 3 |
| num_outlinks_to_geolocated_abroad | 9 |
| percent_inlinks_from_geolocated_abroad | 0.0213 |
| percent_outlinks_to_geolocated_abroad | 0.0122 |
| num_inlinks | 141 |
| num_outlinks | 739 |
| num_bytes | 13815 |
| num_references | 16 |
| num_edits | 471 |
| num_editors | 268 |
| num_discussions | 16 |
| num_pageviews | 639 |
| num_wdproperty | 16 |
| num_interwiki | 59 |
| featured_article | |

# Applications

Leaving user consumption aside, Wikipedia data nurtures very different applications that range from scientific studies to practical tools. By opening the dataset, we want to widen these possibilities, as it includes fine-grained data for all articles in all Wikipedia languages, containing a) computed relevant features (number of edits, number of pageviews, number of Bytes, etc.), and b) all the context related features described above, going from the geolocation and ISO codes, to the flag that determines the final CCC selection. Then, the uses of the dataset are several but we want to highlight three: 1) Wikipedia Culture Gap assessment and improvement, 2) Academic research in the Digital Humanities field, and 3) User-generated Content based technologies.

## Culture Gap Analysis and Improvement

One of the two strategic goals set by the Wikimedia Foundation for the 2030 horizon is to "counteract structural inequalities to ensure a just representation of knowledge and people in the Wikimedia movement"[8]. Hence, fighting to reduce the culture gap between language editions is one the main activities that may benefit from the dataset, which presents a detailed cartography of cultural diversity within each Wikipedia language edition.

The importance of explaining the gap is not only a matter of depicting absolute figures, but of building the capacity to assist editors in discovering valuable articles from other cultural contexts (especially those at a far distance), and in establishing routines to increase the cultural diversity of their language editions. In this sense, we have created lists of 500 articles from every CCC which contain articles that can be considered valuable according to different topical and relevance criteria. Currently, there exist more than ten 'Top CCC article lists', which are created ranking articles by simple or compound relevance characteristics (e.g. number of editors, number of pageviews, creation_date, etc.) and topical subsegments of CCC (e.g. women biographies, geolocated articles, etc.). These selections are presented on tables with all the relevance features for each article along with its availability in a target language (a link to the corresponding article in the target language is provided, or a red label in case it is not existing), so the editor can easily see which articles are valuable and whether they are missing or not in her language in order to create them (Figure 2). General overview tables showing the coverage of Top CCC article lists from all language editions are also provided to editors so they can easily see and start bridging the gap for the languages whose contextual content is mostly missing.

---

[8] https://meta.wikimedia.org/wiki/Strategy/ Wikimedia_movement/2017/Direction

*Figure 2. Top CCC articles in Romanian CCC by number of editors and their availability in Polish Wikipedia*

In addition to these lists, we are currently working on providing different culture gap analysis for specific periods of time with data visualizations and a newsletter[9]. We believe showing on a monthly basis how many articles are being created for each CCC (and for each country) in each language may motivate and facilitate editors to correct the gap and incorporate some new editing routines.

Much more can be done in this direction starting from the Wikipedia Cultural Diversity dataset. For example, more complex criteria could be defined for ranking CCC article from a given language, such as composite criteria to highlight the articles that are more central within the cultural context for a given language, or that are more exclusive or more representative of that context.

**Academic Research on Digital Humanities**

Even though not all people from all cultures, languages and territories are able to contribute to Wikipedia, the online encyclopedia arguably represents the most complete picture on the world cultural diversity knowledge in the Internet. Hence, this dataset allows to study and compare knowledge representation from different linguistic communities and may foster all kinds of academic intersections between Humanities in the digital era.

The CCC set of articles from a language edition should not necessarily be taken as a whole, as they can be split by category, and specific topics can be object of a more focused analysis. For instance, the study of the creation and attention dedicated to CCC or to specific parts of it is another avenue of research to explain the informational needs of a community. In this sense, we observed that articles in CCC tend to be much more developed (in number of bytes, references, images, etc.), and also gather more attention (in number of pageviews) than the rest of Wikipedia articles. In other words, CCC typically represents a specially important part of a Wikipedia, where editors invest more effort, and likewise consulting these articles ends up being a central use of the encyclopedia by readers (Miquel-Ribé, 2016). Another interesting research direction is the study of overlaps and mutual coverage between CCC from different languages to unveil the relationship between linguistic communities and their representation of knowledge.

**User-generated Content based technologies**

Wikipedia is the largest free knowledge repository in many languages, and so it is used to feed many applications that use technologies such as Natural Language Processing (NLP) or information retrieval (IR) (Gabrilovich & Markovitch, 2009; Han, 2014). The Wikipedia Cultural Diversity dataset can feed service algorithms as it contains characteristics in the dimensions of relevance and localness, both suitable in order to tailor for example the results of a search engine or a social media news feed. Considering that many computer systems utilize user-generated content, offering them a categorized and enriched version of Wikipedia metadata may allow for improvements and better personalization.

## Conclusions and Future Work

With the dataset presented in this paper we expect to remove some of the main impediments to both recognize and foster cultural diversity in Wikipedia, as well as to stimulate cross-cultural research in the field of Digital Humanities. These are the most important aims for the Wikipedia Cultural Diversity project[10]. The dataset is made available for the 300 language editions, and contains a fine-grained categorization of each article's relationships towards their nearby geographical and cultural entities, that enable insights into how different linguistic communities define themselves.

---

[9] https://meta.wikimedia.org/wiki/Grants:Project/WCDO/Culture_Gap_Monthly_Monitoring

[10] https://meta.wikimedia.org/wiki/Wikipedia_Cultural_Diversity_Observatory

Most importantly, the Wikipedia Cultural Diversity dataset represents a complete cartography of cultural diversity in Wikipedia, which allows editors to gauge the knowledge gaps between language editions. This has a vital importance when it comes to developing strategies in order to bridge such gaps.

The dataset released includes all the features employed by the classifier, which constitute a valuable enrichment of the metadata extracted from the Wikimedia databases. All the code used to process the data and to create the dataset is also released, as well as the results of the manual assessment[11]. The high degree of agreement between the algorithm and the human raters indicates the reliability of the classification.

In the foreseeable future we aim to extend this work in two particular areas: a) the complete automatization of the Wikipedia Cultural Diversity dataset creation on a monthly basis, and b) the enrichment of the dataset with new features. For example, it would be interesting to define and compute some metric to measure centrality of an article within CCC, or other indicators of the relevance of an article within a specific cultural context. Other features based on the text and elements such as the images could also add further nuances to the article characterization. Current and future versions of the Wikipedia Cultural Diversity dataset can be used to understand better the findings from previous Wikipedia scholarly studies as well as they can trigger new avenues for research.

## Acknowledgements

We thank Andreas Kaltenbrunner for his valuable feedback on the methodology, Bora Edizel and Diego Saez-Trumper for their help on defining the machine learning algorithm.
This work was partially funded by a Project Grant from Wikimedia Foundation and supported by the Catalan Agency for Business Competitiveness, ACCIÓ, under the AlgoFair Project.

## References

Gabrilovich, E., & Markovitch, S. 2009. *Wikipedia-based semantic interpretation for natural language processing. Journal of Artificial Intelligence Research*.

Dyer, C. 2014. *Notes on Noise Contrastive Estimation and Negative Sampling. CoRR Abs/1312.0976, 1410*, arXiv:1410.8251.

Hale, S. A. 2014. *Multilinguals and Wikipedia editing*. Presented at the WS '14: Proceedings of the 2014 ACM conference on Web science.

Han, B. 2014. *Improving the Utility of Social Media with Natural Language Processing*, 1–232.

Hecht, B. J., & Gergle, D. 2010a. *On the localness of user-generated content*. Presented at the CSCW '10: Proceedings of the 2010 ACM conference on Computer supported cooperative work.

Hecht, B., & Gergle, D. 2010b. *The tower of Babel meets web 2.0: user-generated content and its applications in a multilingual context*. Presented at the CHI '10: Proceedings of the SIGCHI Conference on Human Factors in Computing Systems, New York, New York, USA: ACM Request Permissions.

Kim, S., Park, S., Hale, S. A., Kim, S., Byun, J., & Oh, A. H. 2016. *Understanding Editing Behaviors in Multilingual Wikipedia. PloS One*, *11*(5), e0155305.

Lemmerich, F., Sáez-Trumper, D., West, R., & Zia, L. 2018. Why the World Reads Wikipedia: Beyond English Speakers. *arXiv preprint arXiv:1812.00474*.

Mesgari, M., Okoli, C., Mehdi, M., Nielsen, F. Å., & Lanamäki, A. 2015. *"The sum of all human knowledge": A systematic review of scholarly research on the content of Wikipedia*. Journal of the Association for Information Science and Technology, *66*(2), 219–245.

Miquel-Ribé, M. 2016. *Identity-based Motivation in Digital Engagement: The Influence of Community and Cultural Identity on Participation in Wikipedia*. Doctoral dissertation. Universitat Pompeu Fabra.

Miquel-Ribé, M., & Laniado, D. 2016. Cultural Identities in Wikipedias (pp. 24–10). Proceedings of the 7th 2016 International Conference on Social Media & Society, New York, New York, USA: ACM.

Miquel-Ribé, M. & Laniado, D. 2018. Wikipedia Culture Gap: Quantifying Content Imbalances Across 40 Language Editions. Frontiers in Physics, 6(1), 54.

Okoli, C. 2014. *Wikipedia in the eyes of its beholders: A systematic review of scholarly research on Wikipedia readers and readership.* Journal of the Association for Information Science and Technology, 65(12), 2381-2403.

Okoli, C., Mehdi, M., Mesgari, M., Nielsen, F. Å., & Lanamäki, A. 2012. *The People's Encyclopedia Under the Gaze of the Sages: A Systematic Review of Scholarly Research on Wikipedia*. SSRN Electronic Journal.

Samoilenko, A., Karimi, F., Edler, D., Kunegis, J., & Strohmaier, M. 2016. *Linguistic neighbourhoods: explaining cultural borders on Wikipedia through multilingual co-editing activity*. EPJ Data Science, *5*(1), 171–21.

Schroeder, R., & Taylor, L. 2015. *Big data and Wikipedia research: social science knowledge across disciplinary divides*. Information, Communication & Society, 18(9), 1039–1056.

Warncke-Wang, M., Uduwage, A., Dong, Z., & Riedl, J. 2012. *In search of the ur-Wikipedia: universality, similarity, and translation in the Wikipedia inter-language link network.* OpenSym '12: Proceedings of the Eighth Annual International Symposium on Wikis and Open Collaboration, 20.

---

[11] http://github.com/marcmiquel/wcdo